\documentclass[twocolumn,showpacs,preprintnumbers,amsmath,amssymb]{revtex4}

\usepackage{graphicx}
\usepackage{dcolumn}
\usepackage{bm}

\newcommand{\R}{{\mathbb R}}

\newcommand{\Z}{{\mathbb Z}}
\newcommand{\C}{{\mathbb C}}

\newcommand{\be}{\begin{equation}}
\newcommand{\ee}{\end{equation}}
\newcommand{\bea}{\begin{eqnarray}}
\newcommand{\eea}{\end{eqnarray}}
\newcommand{\ul}{\underline}
\newcommand{\ti}{\tilde}

\newcommand{\spr}[2]{\langle #1 , #2 \rangle}
\newcommand{\id}{\mathbb{I}}
\newcommand{\I}{\mathrm{i}}

\newcommand{\ulz}{\underline{z}}
\newcommand{\di}{\mathcal{D}}
\newcommand{\vrc}{\ul{\Xi}_{E_0}}

\newcommand{\Rg}[1]{R_{2g+2}^{1/2}(#1)}

\begin{document}


\title[Stability of Periodic Soliton Equations]{Stability of Periodic Soliton Equations under Short Range Perturbations}

\author{Spyridon Kamvissis}
 \email{spyros@tem.uoc.gr}
 \affiliation{Department of Applied Mathematics, University of Crete \\
714 09 Iraklion, Greece}
\author{Gerald Teschl}%
 \email{Gerald.Teschl@univie.ac.at}
 \homepage{http://www.mat.univie.ac.at/~gerald/}%
 \affiliation{Faculty of Mathematics, University of Vienna,
Nordbergstrasse 15, 1090 Vienna, Austria\\ and\\ International Erwin Schr\"odinger
Institute for Mathematical Physics, Boltzmanngasse 9, 1090 Vienna, Austria}

\date{\today}
\thanks{Research supported in part by the ESF programme MISGAM, and
the Austrian Science Fund (FWF) under Grant No.\ P17762.}

\begin{abstract}
We consider the stability of (quasi-)periodic solutions of soliton equations under short range
perturbations and give a complete description of the related long time asymptotics.

So far, it is generally  believed that a perturbed periodic
integrable system  splits asymptotically into a number of solitons plus a decaying
radiation part, a situation similar to that observed for perturbations of the constant solution.
We show here that this
is not the case; instead  the radiation part does not decay, but
manifests itself  asymptotically as a modulation of the periodic
solution which undergoes a continuous phase transition in the isospectral class of
the periodic background solution.

We provide an explicit formula for this modulated solution in terms of Abelian integrals 
on the underlying hyperelliptic Riemann surface and provide numerical evidence for its
validity. We use the Toda lattice as a model but the same methods and ideas
are applicable to all soliton equations in one space dimension (e.g.\ the Korteweg-de Vries
equation).
\end{abstract}


\maketitle

\section{Introduction}

One of the most important defining properties of solitons  is their
stability under perturbations. The classical result going back to
Zabusky and Kruskal \cite{zakr} states that a "short range" perturbation
of the constant solution of a soliton equation
eventually splits into a number of  stable solitons and a decaying background
radiation component \cite{km}. So the solitons constitute
the stable part of arbitrary short range initial conditions. It is
generally believed, and is claimed in \cite{kumi}, that this remains valid
when the constant background solution is replaced by a quasi-periodic one. 
Our aim here is to show this is not the case, and that a thus far undiscovered
phenomenon appears in the description of the long time asymptotics.

Solitons on a (quasi-)periodic background have a long tradition
and are used to model localized excitements on a phonon, lattice, or magnetic
field background. Consequently, periodic solutions, as well as solitons travelling
on a periodic background, are well understood. The first results were given
over thirty years ago in the pioneering work by Kuznetsov and Mikha\u\i lov
\cite{kumi}, where the stability of solitons of the
Korteweg-de Vries equation on the background of the two-gap Weierstrass
solution is investigated. There the $N$-soliton solution on this
background is computed and it is shown that each soliton produces a phase shift.
In addition, it is claimed that the asymptotic state of any short range
initial condition is a set of solitons.

However, we will show that the asymptotic state is more complicated. The reason for
this is related to the fact that the phase shifts of the solitons do not necessarily add up
to zero. Hence there must be an additional feature making up for the overall phase shift.
One might conjecture that solitons alone suffice to describe the asymptotic
state at least in the case where the phase shifts add up to zero --- which is the
situation assumed in \cite{kumi}. However, we will show that even if no solitons are
present, the asymptotic state is not just the periodic background.

Due to the lack of powerful asymptotic analysis tools and computer power at the time
of \cite{kumi} it was impossible to identify this contribution then. However, it seems
that this omission was neither pointed out in the literature nor was a complete description
of the asymptotic state known.

To illustrate these facts, let us consider the doubly infinite Toda lattice in Flaschka's
variables (see e.g.\ \cite{tjac} or \cite{ta})
\begin{align*}
\dot b(n,t) &= 2(a(n,t)^2 -a(n-1,t)^2),\\
\dot a(n,t) &= a(n,t) (b(n+1,t) -b(n,t)),
\end{align*}
$(n,t) \in \Z \times \R$,
where the dot denotes differentiation with respect to time. We
will consider a quasi-periodic algebro-geometric background
solution $(a_q,b_q)$ (e.g., any periodic solution) plus
a short range (in the sense of \cite{emt2})
perturbation $(a,b)$. The
perturbed solution can be computed via the inverse
scattering transform. The case where $(a_q,b_q)$ is constant
is classical (see again \cite{tjac} or \cite{ta}), but the more
general case applicable  here has  only recently been
analyzed in \cite{emt2}
(see also \cite{emt}).

\begin{figure}
\includegraphics[width=7cm]{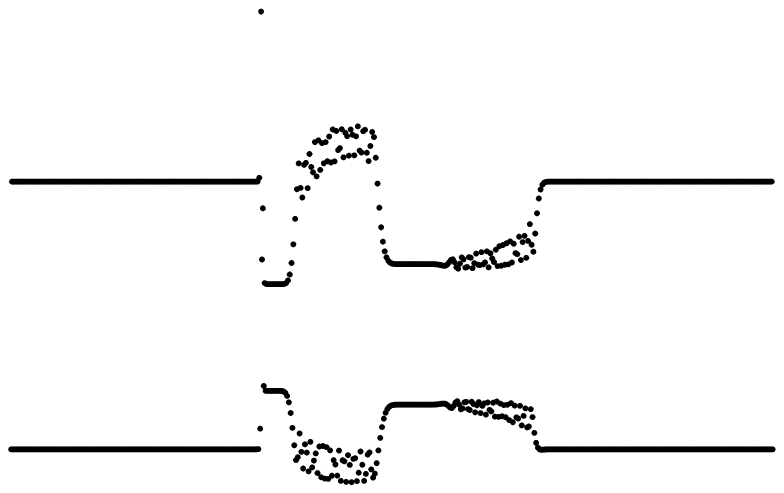}\\
\includegraphics[width=7cm]{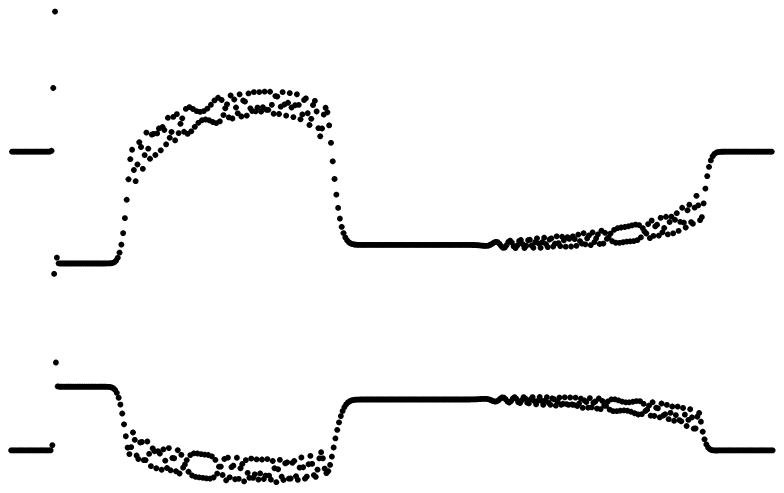}
\caption{Numerically computed solution of the Toda lattice, with initial
condition a period two solution perturbed at one point in the middle.} \label{fig1}
\end{figure}

In Figures~\ref{fig1} the numerically computed solution corresponding to
the initial condition $a(n,0)=1$, $b(n,0)=(-1)^n + 2 \delta_{0,n}$ is shown.
In each of 
these two pictures the two observed lines express the variables $a(n,t)$ at two frozen times $t=90$
and $t=165$.
In areas where the lines seem to be continuous this is due to the fact that we have plotted
a huge number of particles (around $1000$) and also due to the $2$-periodicity in space.
So one can think of the two lines as the even- and odd-numbered particles of the lattice.
We first note the single perturbation which separates two regions of
apparent periodicity on the left. Comparing the two pictures we note that
this is a solitary wave travelling to the left. Our computations
show that it is indeed a soliton. To the right of the soliton, we observe three different 
areas of apparent periodicity (with period two).
Between these areas there are  transitional regions 
which interpolate between the different period two regions. 

It is the purpose of this paper to give a complete explanation of
this picture. We provide the details in case of the Toda lattice though it is
clear that our arguments apply to other soliton equations as well.

\section{Quasiperiodic solutions in terms of Riemann theta functions}

We begin by recalling that quasi-periodic solutions are most conveniently written
in terms of Riemann theta functions (\cite{dub}). Such explicit formulae were 
first derived in the mid-70s thanks to the pioneering work of
the Soviet school (Novikov, Its, Matveev, etc.).
In the particular case of the Toda lattice the underlying Riemann surface is associated with the
square root
\[
\Rg{z} = -\prod_{j=0}^{2g+1} \sqrt{z-E_j}, 
\]
where the numbers $E_j\in\R$ are the band edges of the spectrum $\Sigma =
\bigcup_{j=0}^g [E_{2j},E_{2j+1}]$ of the corresponding
Jacobi operator (see \cite{tjac}).
We can picture this surface as  the end result of cutting and pasting; we consider two 
"sheets" (copies of the complex plane), we cut them along the segments
$[E_0,E_1]$, $[E_2, E_3]$, \dots{} and paste the top side of the upper sheet
to the bottom side of the lower sheet and the bottom side of the upper
sheet to the top side of the lower sheet across every such segment.

On this Riemann surface we have a standard
basis of normalized holomorphic differentials
\[
\zeta_j = \frac{\sum_{k} c_j(k) z^k}{\Rg{z}}dz, \quad1\le j \le g
\]
(the constants $c_j(k)$ have to be determined from the usual normalization
with respect to a canonical homology basis).

Introduce the vector
\begin{align*}
z_j(n,t) &= \int_{E_0}^{\infty_+} \zeta_j - \sum_{k=1}^g \int_{E_0}^{\mu_k} \zeta_j\\
& - n \int_{\infty_-}^{\infty_+} \zeta_j + t 2 c_j(g) - \vrc \in \C^g.
\end{align*}
Here $\infty_\pm$ are the points above $\infty$ on the upper/lower sheet and
$\vrc$ is the vector of Riemann constants. The numbers $\mu_j$ are some arbitrary points whose images in the complex plane lie in the $j$-th interior spectral gap.
All possible choices form the isospectral class of quasi-periodic Jacobi operators
with the given spectral bands.  This isospectral class is just a $g$ dimensional torus,
since the preimage of each gap with respect to the map that maps the surface to 
the complex plane (seen as one of the two sheets) consists of two parts,
one on the upper and one the lower sheet, which form a circle.

Then the well-known formulas for the solutions read
\begin{align*}
a(n,t)^2 &= \ti{a}^2 \frac{\theta(\ulz(n+1,t))
\theta(\ulz(n-1,t))}{\theta(\ulz(n,t))^2},\\
b(n,t) &= \ti{b} + \frac{1}{2}\frac{d}{dt}
\ln\left( \frac{\theta(\ulz(n,t))}{\theta(\ulz(n-1,t))} \right),
\end{align*}
where $\ti{a}$, $\ti{b}$ are the averages and
\[
\theta(\ulz) = \sum_{\ul{m} \in \Z^g} \exp 2 \pi \I \left( \spr{\ul{m}}{\ulz} +
\frac{\spr{\ul{m}}{\ul{\tau} \, \ul{m}}}{2}\right) ,\quad \ulz \in \C^g,
\]
is the Riemann theta function of our surface. The matrix $\tau$ is the matrix of
$b$-periods of the normalized basis of holomorphic differentials $\zeta_j$.

\section{The main result}

We will assume for simplicity that no solitons are present.
In other words, we assume that the Jacobi operator
above has no eigenvalues. The assumption is not crucial, since the solitons
can be easily incorporated using a Darboux transform.

To obtain the long time asymptotics one reformulates the problem
as a Riemann-Hilbert problem (RHP) on the underlying Riemann surface.
This shows that the solution can be read off from the two by two matrix valued
function $m$ that is meromorphic off the preimage of the spectrum $\Sigma$,
with divisor satisfying
\[
(m_{j1}) \ge -\sum_k \di_{\mu_k^*(n,t)}, \quad
(m_{j2}) \ge -\sum_k \di_{\mu_k(n,t)},
\]
$ j=1,2$, jump matrix given by
\[
J(p,n,t)= \begin{pmatrix}
1 - |R(p,n,t)|^2 & - \overline{R(p,n,t)} \\
R(p,n,t) & 1
\end{pmatrix},
\]
$p\in\Sigma$, and normalization
\[
m(p,n,t) \to \id, \quad \text{as } p\to \infty_+.
\]
This is  similar to the RHP applicable to  the constant background case, 
with the main difference being that we allow poles at the points
$\mu_j(n,t)$, and their flip images $\mu_j^*(n,t)$ on the other sheet.
The $g$ points $\mu_j(n,t)$ are uniquely defined by the Jacobi inversion problem
\[
\sum_{k=1}^g \int_{E_0}^{\mu_k(n,t)} \zeta_j= 
\sum_{k=1}^g \int_{E_0}^{\mu_k} \zeta_j\\
+ n \int_{\infty_-}^{\infty_+} \zeta_j - t 2 c_j(g).
\]
This is a crucial point and related to the fact that our RHP is no
longer formulated in the complex plane. While a holomorphic RHP with
jump of index zero has a solution in the complex plane, this
is no longer true on a surface of genus $g$ unless we admit
at least $g$ poles (\cite{ro}). In particular, the above RHP
has no holomorphic solutions except in special cases
(e.g.\ if there is no jump). In fact, a main issue in the mathematical
analysis of the RHP (\cite{katept}) is to define an appropriate space of solutions
which makes the problem well-posed.

The matrix elements of the jump are of the form
\[
R(p,n,t) = R(p) \Theta(p,n,t) \exp( t \phi(p) ),
\]
where $R(p)$ is the reflection coefficient at $t=0$, $\Theta$ is a ratio of four
theta functions
\[
\Theta(p,n,t) = \frac{\theta(\ulz(p,n,t))}{\theta(\ulz(p,0,0))}
\frac{\theta(\ulz(p^*,0,0))}{\theta(\ulz(p^*,n,t))}
\]
(of modulus one), and the phase $\phi$ is
given by
\[
\phi(p) = 2 \int_{E_0}^p \Omega_0 + 2 \frac{n}{t} \int_{E_0}^p \omega_{\infty_+\infty_-}.
\]
Here
\[
\omega_{\infty_+ \infty_-}= \frac{\prod_{j=1}^g (z -\lambda_j) }{\Rg{z}}dz
\]
is the normalized Abelian differential of the third kind with poles
at $\infty_+$ respectively $\infty_-$ and
\[
\Omega_0 = \frac{\prod_{j=0}^g (z - \ti\lambda_j) }{\Rg{z}}dz,
\quad \sum_{j=0}^g \ti\lambda_j = \frac{1}{2} \sum_{j=0}^{2g+1} E_j,
\]
is the normalized Abelian differential of the second kind with second order poles at
$\infty_+$ respectively $\infty_-$. The constants $\lambda_j$ respectively
$\ti\lambda_j$ again have to be determined from the normalization.

There are $g+1$ stationary phase points $z_j(n/t)$ (roots of
$\phi'$) which behave as follows:
As $\eta=\frac{n}{t}$ runs from $-\infty$ to $+\infty$ we start with $z_g(\eta)$
moving from $\infty$ towards $E_{2g+1}$ while the others stay in their
spectral gaps until $z_g(\eta)$ has passed the first spectral band.
After this has happened, $z_{g-1}$ can leave its gap, while $z_g(\eta)$
remains there, traverses the next spectral band and so on. Until
finally $z_0(\eta)$ traverses the last spectral band and escapes to
$-\infty$.

Factorizing the jump matrix and using the asymptotic  analysis of oscillatory
Riemann-Hilbert problems introduced in (\cite{dz}), but generalized accordingly
in \cite{katept}, we have shown  that for long times the perturbed
Toda lattice is asymptotically close to the 
following limiting lattice defined by
\[
\aligned
\prod_{j=n}^{\infty} (\frac{a_l(j,t)}{a_q(j,t)} )^2 = &
\frac{\theta(\ulz(n,t))}{\theta(\ulz(n+1,t))}
\frac{\theta(\ulz(n+1,t)+ \ul{\delta}(n,t))}{\theta(\ulz(n,t)+\ul{\delta}(n,t))} \times\\
& \times \exp\left( \frac{1}{2\pi\I} \int_{C(n/t)} 
\hspace*{-7mm} \log (1-|R|^2) \omega_{\infty_- \infty_+}\right),\\
\delta_\ell(n,t)= & \frac{1}{2\pi\I} \int_{C(n/t)}
\hspace*{-7mm} \log (1-|R|^2) \zeta_\ell,
\endaligned
\]
where $R$ is the associated reflection coefficient,
$C(n/t) = \Sigma \cap (-\infty,z_j(n/t))$, and
$z_j(n/t)$ is the single stationary phase
point lying in the spectrum, if there is such a point, or
otherwise, one of the two stationary phase points 
lying in the same spectral gap.

In other words, as in \cite{dz} and \cite{km} one has to factorize the jump matrix
and deform the contour according to the stationary phase points. The deformed
RHP is asymptotically close to the unperturbed one by the oscillatory nature of
the jump. However, due to the poles at $\mu_j(n,t)$ there is an additional
contribution which gives raise to the modulated limiting lattice. We thus have here
an extension of the so called nonlinear stationary phase method to problems defined
on a Riemann surface.

In summary, for any short range perturbation $a(n,t)$ of a quasi-periodic solution of the
Toda lattice one has that $\prod_{j=n}^{\infty} \frac{a(j,t)}{a_l(j,t)} \to 1$
uniformly in $n$, as $t \to \infty$. From this one recovers the $a(n,t)$ and by differentiating
one recovers the $b(n,t)$. It thus follows that
\[
|a(n,t) - a_l(n,t)| + |b(n,t)- b_l(n,t)| \to 0
\]
uniformly in $n$, as $t \to \infty$.

If solitons are present we can apply appropriate Darboux transformations
to add the effect of such solitons. What we then see asymptotically is
additional travelling solitons on a periodic background (\cite{emt3} or \cite{km2}).

\begin{figure}
\includegraphics[width=7cm]{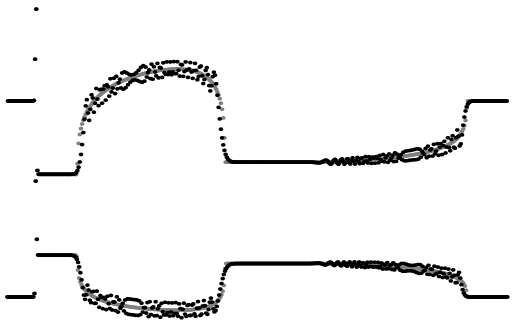}
\caption{Solution of the Toda lattice (black) plus corresponding limiting lattice (gray).
No gray points visible implies that black points overlap gray points.} \label{fig2}
\end{figure}

We are not presenting here our detailed computations leading to the asymptotic
identification of the perturbed problem described in the introduction
and the modulated lattice described above.
A complete mathematical proof will be given in \cite{katept}.
Instead, we are here providing a numerical confirmation of the above result.
Indeed the limiting lattice can be easily computed
numerically. For the initial data defined in  the introduction a and at time
$t=90$ the result is shown in
Figure~\ref{fig2}.
The soliton was added using a Darboux transformation.

\section{Conclusion}

Let $g$ be the genus of the hyperelliptic Riemann surface associated with
the unperturbed solution. We have shown that the $n/t$-plane contains $g+2$ areas where
the perturbed solution is close to a quasi-periodic solution in the same isospectral
torus. In between there are $g+1$ regions where the perturbed solution is asymptotically
close to a modulated lattice which undergoes a
continuous phase transition (in the Jacobian variety) and which interpolates
between these isospectral solutions. In the special case of  the free solution ($g=0$) the
isospectral torus consists of just one point and we recover the classical result.

To conclude, we believe that, apart from the interesting mathematical problem of the
analysis of Riemann-Hilbert problems
on Riemann surfaces, the limiting lattice is interesting on its own.
We even speculate that there may even be a physical or technological interest in such
a modulated lattice and that the asymptotic problem described here might provide
a feasible way of producing such modulated solutions.

\begin{acknowledgments}
Spyridon Kamvissis gratefully acknowledges the support of the European Science
Foundation (ESF) and the kind hospitality of the Faculty of Mathematics
during two visits to the University of Vienna from April to July 2005
and in April 2006.
\end{acknowledgments}

\end{document}